\documentclass[seceq]{ptptex}
\usepackage{graphicx}
\title{%        %You can use \\ for explicit line-break
Survival of charmonia above $T_c$ in anisotropic lattice QCD%
}

\author{%       %Use \scshape  for the family name
Hideaki \textsc{Iida}$^{1,}$\footnote{E-mail: iida@yukawa.kyoto-u.ac.jp}, 
Takumi \textsc{Doi}$^{2}$, 
Noriyoshi \textsc{Ishii}$^{3}$, 
Hideo \textsc{Suganuma}$^{4}$ \\ and 
Kyosuke~\textsc{Tsumura}$^{5}$%
}

\inst{%     %Affiliation, neglected when [addenda] or [errata]
$^1$YITP, Kyoto University, Sakyo, Kyoto 606-8502, Japan\\
$^2$ Dept. of Phys. \& Astr., University of Kentucky, Lexington KY 40506, USA\\
$^3$CCS, University of Tsukuba, Tsukuba, Ibaraki~305-8577, Japan\\
$^4$Department of Physics, Kyoto University, Sakyo, Kyoto 606-8502, Japan\\
$^5$Analysis Technology Center, Fujifilm Corporation, Kanagawa 250-0193, Japan
}

\abst{
We find a strong evidence for the survival of $J/\Psi$ and $\eta_c$ 
as spatially-localized $c\bar c$ (quasi-)bound states above the QCD critical temperature $T_c$, 
by investigating the boundary-condition dependence of their energies and spectral functions. 
In a finite-volume box, there arises a boundary-condition dependence 
for spatially spread states, while no such dependence appears for spatially compact states.
In lattice QCD, we find almost {\it no} spatial boundary-condition dependence for the 
energy of the $c\bar c$ system in $J/\Psi$ and $\eta_c$ channels for $T\simeq(1.11-2.07)T_c$. 
We also investigate the spectral function of charmonia above $T_c$ in lattice QCD 
using the maximum entropy method (MEM) in terms of the boundary-condition dependence.
There is {\it no} spatial boundary-condition dependence for the low-lying peaks corresponding to 
$J/\Psi$ and $\eta_c$ around 3GeV at $1.62T_c$. These facts indicate the survival of 
$J/\Psi$ and $\eta_c$ as compact $c\bar c$ (quasi-)bound states for $T_c < T < 2T_c$.} 

\begin{document}

\maketitle

\section{Introduction}

$J/\Psi$ suppression \cite{HMHK86,MS86} was theoretically proposed as an important signal 
of quark-gluon plasma (QGP). The basic idea of $J/\Psi$ suppression is that 
$J/\Psi$ disappears above the QCD critical temperature $T_c$ due to vanishing of the 
confinement potential and appearance of the Debye screening effect. 
In contrast, recently, some lattice QCD calculations indicate an opposite result that $J/\Psi$ 
and $\eta_c$ survive even above $T_c$ \cite{UKMM01,UNM02,AH04,DKPW04,IDIST06,AAOPS07}. 
%In Ref.~\cite{UKMM01}, the authors find the strong spatial correlation between $c$ and $\bar c$. 
Spectral functions of charmonia are extracted from temporal correlators at high temperature 
using the maximum entropy method (MEM) in Refs.~4)-8). Although there are 
some quantitative differences, the peaks corresponding to $J/\Psi$ and $\eta_c$ seem to 
survive even above $T_c$ ($T_c < T < 2T_c$) in the $c\bar c$ spectral function. 

However, one may ask a question on the ``survival of $J/\Psi$ and $\eta_c$'' 
observed in lattice QCD. Are the $c\bar c$ states above $T_c$ observed in lattice QCD 
really compact (quasi-)bound states? 
Since colored states are allowed in the QGP phase, there is a possibility that the observed 
$c\bar c$ state in lattice QCD is just a $c\bar c$ scattering state,\cite{MP08} 
which is spatially spread. 
Particularly in lattice QCD, even scattering states have discretized spectrum, 
due to the finite-volume effect. 
Therefore, for the fair judgment of the ``survival of charmonia above $T_c$", 
it is necessary to clarify whether the $c\bar c$ systems are spatially compact 
(quasi-)bound states or scattering states.

To this end, we study the spatial {\it boundary-condition dependence} of 
the energy and the spectral function for the $c\bar c$ systems ($J/\Psi$ and $\eta_c$) 
above $T_c$ in lattice QCD\cite{IDIST06}. 
Actually, for $c\bar c$ scattering states, there occurs a significant energy difference 
between periodic and anti-periodic boundary conditions on the finite-volume lattice. 
In contrast, for spatially compact charmonia, there is almost no energy difference 
between these boundary conditions.\cite{IDIST06} Using this fact, we investigate the 
$c\bar c$ system above $T_c$ in terms of their spatial extent. 

\section{Boundary-Condition Dependence and Compactness of States}
We show briefly the method to distinguish spatially-localized states from scattering states 
on the finite-volume lattice.\cite{IDIST06,IDIOOS05} For a compact $c\bar c$ (quasi-)bound state, 
the wave-function of the $c\bar c$ state is spatially localized, 
and therefore its energy is insensitive to the spatial boundary condition. 
In contrast, for a $c\bar c$ scattering state, the wave-function of the $c\bar c$ system is 
spatially spread, so that there occurs a significant boundary-condition dependence of 
the energy for the low-lying $c\bar c$ scattering state. 
Let us estimate the boundary-condition dependence. 
\begin{itemize}
\item
Under the {\it periodic boundary condition (PBC)}, the momentum of a quark or an anti-quark is 
discretized as $p_k=2n_k\pi/L \ (k=1,2,3, \ n_k\in {\bf Z})$ on the finite lattice 
with the spatial volume $L^3$, and the minimum momentum is $\vec p_{\rm min}=\vec 0$. 
\item
Under the {\it anti-periodic boundary condition (APBC)}, the momentum is discretized 
as $p_k=(2n_k+1)\pi/L \ (k=1,2,3, \ n_k\in {\bf Z})$. In this case, the minimum momentum 
is $|\vec p_{\rm min}|= \sqrt{3}\pi/L$. 
\end{itemize}
The energy difference of the low-lying $c\bar c$ scattering state is estimated as 
%\begin{eqnarray}
$
\Delta E_{\rm scatt}\equiv E_{\rm APBC}^{\rm scatt}-E_{\rm PBC}^{\rm scatt} 
\simeq 2\sqrt{m_c^2+3\pi^2/L^2} -2m_c \simeq 350{\rm MeV}
$
%\end{eqnarray}
for $L \simeq 1.55{\rm fm}$ in Ref.~7), 
in the non-interacting case with the charm-quark mass $m_c$.
In Ref.~7), we consider possible correction of 
the energy difference from a short-range potential of Yukawa type between a quark and 
an anti-quark, and find that the correction is negligible compared to the energy difference 
$\Delta E_{\rm scatt}\simeq 350{\rm MeV}$ estimated in the non-interacting case.

\section{Pole-Mass Measurements above $T_c$ in Lattice QCD}

First, we perform the standard pole-mass measurement of low-lying $c\bar c$ systems 
at finite temperature in anisotropic quenched lattice QCD\cite{IDIST06}
with the standard plaquette action at $\beta\equiv 2N_c/g^2=6.10$ and 
the renormalized anisotropy $a_s/a_t=4.0$, {\it i.e.}, $a_t=a_s/4 
\simeq (8.12{\rm GeV})^{-1} \simeq 0.024{\rm fm}$ for the spatial and temporal lattice spacing.
The adopted lattice sizes are $16^3\times (14-26)$, which correspond to 
the spatial volume as $L^3 \simeq (1.55{\rm fm})^3$ and the temperature as (1.11$-$2.07)$T_c$. 
We use 999 gauge configurations, picked up every 500 sweeps after the thermalization of 
20,000 sweeps. For quarks, we use $O(a)$-improved Wilson (clover) action on the 
anisotropic lattice \cite{IDIST06}. We adopt the hopping parameter $\kappa=0.112$, 
which reproduces the masses of charmonia at zero temperature. 
To enhance the ground-state overlap, we use a Gaussian spatially-extended operator 
with the extension radius $\rho=0.2{\rm fm}$ in the Coulomb gauge \cite{IDIST06},
which is found to maximize the ground-state overlap. 
The energy of the low-lying $c\bar c$ state is extracted from the temporal correlator of 
the spatially-extended operators, where the total momentum of the system is projected to be zero.

\begin{table}[t]
\begin{center}
\caption{The energy of the $c\bar c$ system in $J/\Psi$ ($J^P=1^{-}$) and $\eta_c$ ($J^P=0^-$) 
channels on PBC and APBC at $\beta=6.10$ at each temperature. 
The superscripts $J/\Psi$ and $\eta_c$ denote quantities of $J/\Psi$ and $\eta_c$, respectively. 
All the statistical errors are smaller than 0.01GeV. 
The energy difference $E_{\rm APBC}-E_{\rm PBC}$ observed in lattice QCD is also added. 
The observed energy difference is very small, compared to the estimated energy difference 
$\Delta E_{\rm scatt}\simeq 350{\rm MeV}$ 
in the case of the low-lying $c\bar c$ scattering states.}

\begin{tabular}{cccccccc}
\hline
\hline
Temperature & $E^{J/\Psi}_{\rm PBC}$  
& $E^{J/\Psi}_{\rm APBC}$ 
 & $E^{J/\Psi}_{\rm APBC}-E^{J/\Psi}_{\rm PBC}$
 \ \ & \ \ $E^{\eta_c}_{\rm PBC}$  
& $E^{\eta_c}_{\rm APBC}$ 
 & $E^{\eta_c}_{\rm APBC}-E^{\eta_c}_{\rm PBC}$
\\
\hline
$1.11T_c$  &3.05GeV & 3.09GeV  &0.04GeV
 \ \ & \ \ 3.03GeV  & 3.02GeV 
  &$-$0.01GeV
\\
$1.32T_c$  &2.95GeV  & 2.98GeV  &0.03GeV 
 \ \ & \ \ 2.99GeV  & 2.98GeV 
 &$-$0.01GeV\\
$1.61T_c$  &2.94GeV  & 2.98GeV 
 &0.04GeV \ \ & \ \ 3.00GeV  & 2.97GeV  &$-$0.03GeV\\
$2.07T_c$  &2.91GeV  & 2.93GeV 
 &0.02GeV \ \ & \ \ 3.01GeV & 3.00GeV  &$-$0.01GeV\\ 
\hline
\end{tabular}
\label{tab1}
\end{center}
\end{table}
%-----------

Table \ref{tab1} shows the boundary-condition dependence of the low-lying $c\bar c$ state energy 
in $J/\Psi$ ($J^P=1^{-}$) and $\eta_c$ ($J^P=0^{-}$) channels at finite temperatures. 
Both in $J/\Psi$ and $\eta_c$ channels, 
the energy difference between PBC and APBC is less than 40MeV, 
which is much smaller than the energy difference $\Delta E_{\rm scatt}$ in  
the case of the low-lying $c\bar c$ scattering states, {\it i.e.},
$|E_{\rm APBC}-E_{\rm PBC}| \ll \Delta E_{\rm scatt}\simeq 350{\rm MeV}$, 
at all measured temperatures.
These results indicate that the observed $c\bar c$ states 
are {\it spatially-localized (quasi-)bound states} as charmonia of $J/\Psi$ and $\eta_c$ 
for $1.11T_c < T < 2.07T_c$.

Here, we comment on a ``constant contribution" to the meson correlator above $T_c$,
provided by ``wrap-around quark propagation", which is pointed out in Ref.~11). 
The $\eta_c$ channel is {\it free from} this extra contribution, 
but the $J/\Psi$ channel suffers from it.
Actually, in contrast to almost no temperature dependence of the $\eta_c$ mass, 
there is a temperature dependence of the $J/\Psi$ mass, 
which may be an artifact due to the constant contribution to the meson correlator \cite{U07}. 
However, the difference of the $J/\Psi$ mass between $1.11T_c$ and $2.07T_c$ is only about 140MeV, 
and this value is rather small compared to the estimated energy difference 
$\Delta E_{\rm scatt}\simeq 350{\rm MeV}$ for $c\bar c$ scattering states.
Therefore, even including this extra effect, our main conclusion is unchanged 
also for $J/\Psi$ above $T_c$.
In fact, $J/\Psi$ and $\eta_c$ survive as spatially-localized (quasi-)bound states 
for $1.11T_c < T < 2.07T_c$.

\section{MEM Analyses for Spectral Functions above $T_c$ in Lattice QCD}

Next, we investigate the boundary-condition dependence of 
the spectral function $A(\omega)$ of the $c \bar c$ system above $T_c$ 
using the maximum entropy method (MEM) in lattice QCD.\cite{IDIST06}
Using MEM, we extract the spectral function $A(\omega)$ from 
the temporal correlator $G(t)$ of the point source and the point sink, 
where the total momentum is projected to be zero. 
In this calculation, we use the Wilson quark action on a fine lattice at $\beta=7.0$, 
{\it i.e.}, $a_t=a_s/4 \simeq (20.2{\rm GeV})^{-1} \simeq 9.75\times 10^{-3}{\rm fm}$.
The adopted lattice size is $20^3\times 46$, which corresponds to the spatial volume 
$L^3\simeq (0.78{\rm fm})^3$ and the temperature $T \simeq 1.62T_c$. 

Figure 2 shows the spectral functions in $J/\Psi$ and $\eta_c$ channels 
on PBC (dotted line) and APBC (solid line). There appear low-lying peaks around 3GeV, 
which correspond to the charmonia of $J/\Psi$ and $\eta_c$. 
We find no spatial boundary-condition dependence for the low-lying peaks, 
which indicate that the $c\bar c$ states corresponding to $J/\Psi$ and $\eta_c$ 
appear as spatially-localized (quasi-)bound states even above $T_c$.\cite{IDIST06}

\section{Summary and Conclusion}

We have studied $J/\Psi$ and $\eta_c$ above $T_c$ in anisotropic lattice QCD to clarify 
whether these states are spatially-localized (quasi-)bound states or $c\bar c$ scattering states. 
As a result, both in $J/\Psi$ and $\eta_c$ channels, we have found almost no spatial 
boundary-condition dependence of the energy of the low-lying $c\bar c$ system 
even on a finite-volume lattice for $(1.11-2.07) T_c$. 
Also in the MEM analysis, we find no spatial boundary-condition dependence of 
the low-lying peaks corresponding to $J/\Psi$ and $\eta_c$ 
in the spectral function at $T \simeq 1.62T_c$. 
These facts indicate that {\it $J/\Psi$ and $\eta_c$ survive in QGP as 
spatially-localized (quasi-)bound states for $T_c < T < 2T_c$.} 

\begin{figure}[t]
\centerline{
\includegraphics[width=5.4cm]
{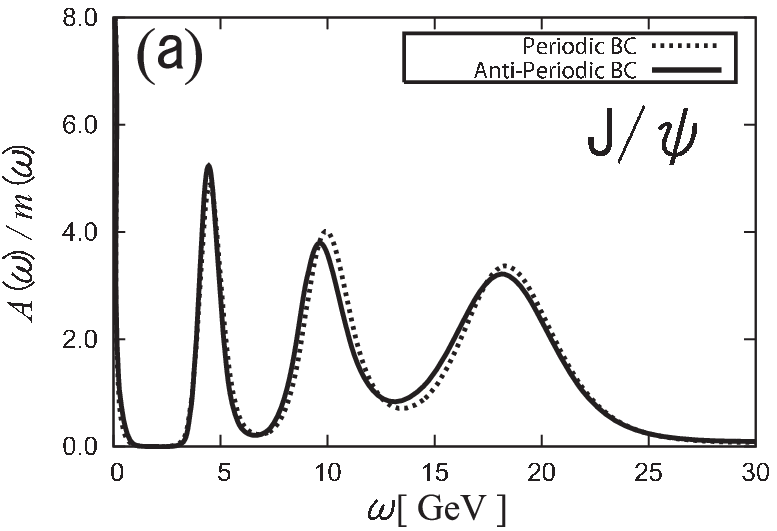}
\hspace{0.5cm}
\includegraphics[width=5.4cm]
{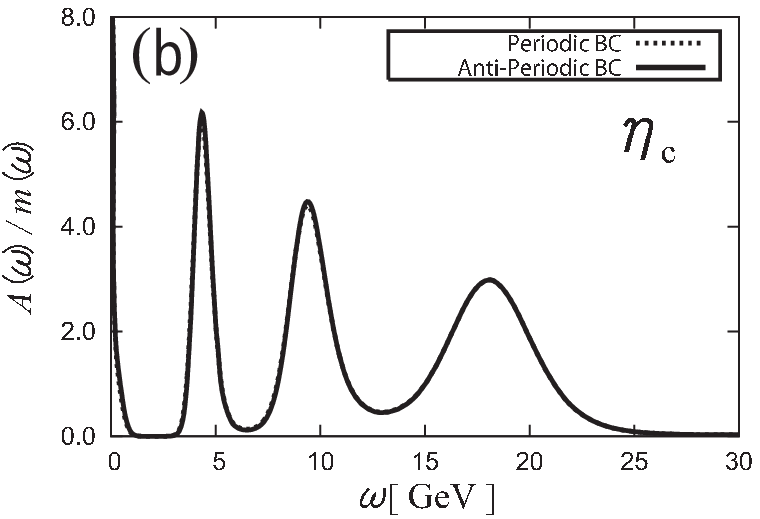}
}
\caption{The spectral function $A(\omega)$ of the $c\bar c$ system  
in (a) $J/\Psi$ channel and (b) $\eta_c$ channel at $1.62T_c$ 
on PBC (dotted line) and APBC (solid line), 
normalized by the default function $m(\omega)$.\cite{IDIST06}
Almost no boundary-condition dependence is found for 
the low-lying peaks around 3GeV, which correspond to the charmonia of $J/\Psi$ and $\eta_c$. 
In the $\eta_c$ channel, the dotted and the solid lines coincide. 
For the figures of the spectral functions with errorbars, see Figs.~9 and 10 in Ref.~7).
}
\label{fig2}
\end{figure}

\section*{Acknowledgement}
H.~I. and H.~S. thank the Yukawa Institute for Theoretical Physics, for fruitful discussions 
at ``New Frontiers in QCD 2008".

%We would like to thank ...........
%The lattice QCD calculation has been done on NEC-SX5 at Osaka University.

%\appendix
%\section{First Appendix} %Empty argument \section{} yields `Appendix'. 
%
%\section{Second Appendix}


\begin{thebibliography}{99}
%%%%%%%%%%%%%%%%%%%%%%%%%%%%%%%%%%%%%%%%%%%%%%%%%%%%%%%%%%%%%
% Some macros are available for the bibliography:
%  o for general use
%    \JL : general journals                 \andvol : Vol (Year) Page
%  o for individual journal 
%    \AJ   : Astrophys. J.           \NC         : Nuovo Cim.
%    \ANN  : Ann. of Phys.           \NPA, \NPB  : Nucl. Phys. [A,B]
%    \CMP  : Commun. Math. Phys.     \PLA, \PLB  : Phys. Lett. [A,B]
%    \IJMP : Int. J. Mod. Phys.      \PRA - \PRE : Phys. Rev. [A-E]     
%    \JHEP : J. High Energy Phys.    \PRL        : Phys. Rev. Lett.
%    \JMP  : J. Math. Phys.          \PRP        : Phys. Rep.
%    \JP   : J. of Phys.             \PTP        : Prog. Theor. Phys.     
%    \JPSJ : J. Phys. Soc. Jpn.      \PTPS       : Prog. Theor. Phys. Suppl.
% Usage:
%  \PRD{45,1990,345}          ==> Phys.~Rev.\ D \textbf{45} (1990), 345
%  \JL{Nature,418,2002,123}   ==> Nature \textbf{418} (2002), 123
%  \andvol{123,1995,1020}    ==> \textbf{123} (1995), 1020
%%%%%%%%%%%%%%%%%%%%%%%%%%%%%%%%%%%%%%%%%%%%%%%%%%%%%%%%%%%%%
  
\bibitem{HMHK86}
T.~Hashimoto, O.~Miyamura, K.~Hirose and T.~Kanki, Phys. Rev. Lett. {\bf 57} (1986), 2123.
\bibitem{MS86}
T.~Matsui and H.~Satz, Phys. Lett. B{\bf 178} (1986), 416.
\bibitem{UKMM01}
T.~Umeda, K.~Katayama, O.~Miyamura and H.~Matsufuru, Int. J. Mod. Phys. A{\bf 16} (2001), 2215; 
H.~Matsufuru, O.~Miyamura, H.~Suganuma and T.~Umeda, AIP Conf. Proc. {\bf CP594} (2001), 258.
\bibitem{UNM02} 
T.~Umeda, K.~Nomura and H.~Matsufuru, hep-lat/0211003, Eur. Phys. J. C{\bf 39} (2005), 9.
\bibitem{AH04} 
M.~Asakawa and T.~Hatsuda, Phys. Rev. Lett. {\bf 92} (2004), 012001. 
\bibitem{DKPW04} 
S.~Datta, F.~Karsch, P.~Petreczky and I.~Wetzorke, Phys. Rev. D{\bf 69} (2004), 094507.
%J.~Phys. G{\bf 31} (2005), S351.
\bibitem{IDIST06} 
H.~Iida, T.~Doi, N.~Ishii, H.~Suganuma and K.~Tsumura, Phys. Rev. D{\bf 74} (2006), 074502; \\
H.~Iida, T.~Doi, N.~Ishii and H.~Suganuma, PoS {\bf LAT2005} (2005), 184.
\bibitem{AAOPS07}
G.~Aarts {\it et al.}
%C.~Allton, M.~Oktay, M.~Peardon and J.~Skullerud, 
Phys. Rev. D{\bf 76} (2007), 094513.
\bibitem{MP08}
$\Acute{{\rm A}}$.~M$\Acute{\rm{o}}$csy and P.~Petreczky, Phys. Rev. D{\bf 77} (2008), 014501.
\bibitem{IDIOOS05}
N.~Ishii, T.~Doi, H.~Iida, M.~Oka, F.~Okiharu and H.~Suganuma, 
Phys. Rev. D{\bf 71} (2005), 034001;
N.~Ishii {\it et al.}, 
%T.~Doi, Y.~Nemoto, M.~Oka and H.~Suganuma, 
Phys. Rev. D{\bf 72} (2005), 074503.
\bibitem{U07}
T.~Umeda, Phys. Rev. D{\bf 75} (2007), 094502.
\end{thebibliography}
\end{document}